\documentclass[12pt,a4paper,reqno]{amsart}
\usepackage[T1]{fontenc}
\usepackage{lmodern}
\usepackage{amsfonts,amssymb,amsmath,mathtools,enumerate}
\usepackage{needspace}
\usepackage[margin=25mm,footskip=10mm]{geometry}
\usepackage[hidelinks]{hyperref}

\newtheorem{theorem}{Theorem}[section]
\newtheorem{proposition}[theorem]{Proposition}
\newtheorem{corollary}[theorem]{Corollary}
\newtheorem{lemma}[theorem]{Lemma}
\theoremstyle{remark}
\newtheorem{remark}[theorem]{Remark}

\numberwithin{equation}{section}
\newcommand{\C}{\mathbb C}
\newcommand{\R}{\mathbb R}
\newcommand{\Z}{\mathbb Z}
\newcommand{\N}{\mathbb N}
\newcommand{\Nzero}{\mathbb N_0}
\newcommand{\As}{A_{s,h}}
\newcommand{\Rs}{R_s}
\newcommand{\Gs}{\Gamma_s}
\newcommand{\E}{\mathcal E}
\newcommand{\Lloc}{L^1_{\mathrm{loc}}}
\newcommand{\Id}{\mathrm{Id}}
\newcommand{\Prob}{\mathbb P}
\DeclareMathOperator{\Dom}{Dom}
\DeclareMathOperator{\End}{End}

\title[Riemann--Liouville operators and generalized Stirling numbers]{Normal
ordering of Riemann--Liouville operators\\ and a probabilistic formula\\
for generalized Stirling numbers}
\author[Sodbaatar Adiya]{Sodbaatar Adiya\textsuperscript{1,2}}
\author[Khongorzul Dorjgotov]{Khongorzul Dorjgotov\textsuperscript{2,3}}
\author[Bayarmagnai Gombodorj]{Bayarmagnai Gombodorj\textsuperscript{2,3}}
\author[Uuganbayar Zunderiya]{Uuganbayar Zunderiya\textsuperscript{2,3}}
\address{\textsuperscript{1}National Intelligence University, Ulaanbaatar, Mongolia}
\address{\textsuperscript{2}Department of Mathematics, National University of Mongolia,\newline
\hspace*{\parindent}Ulaanbaatar, Mongolia}
\address{\textsuperscript{3}Center for Mathematics and Mathematics Education, National University of Mongolia,\newline
\hspace*{\parindent}Ulaanbaatar, Mongolia}
\email[Sodbaatar Adiya]{sodoomath@gmail.com}
\email[Khongorzul Dorjgotov]{pilpalpil@gmail.com}
\email[Bayarmagnai Gombodorj]{bayarmagnai@smcs.num.edu.mn}
\email[Uuganbayar Zunderiya]{zunderiya@gmail.com}
\date{}
\hypersetup{
  pdftitle={Normal ordering of Riemann--Liouville operators and a probabilistic formula for generalized Stirling numbers},
  pdfauthor={Sodbaatar Adiya, Khongorzul Dorjgotov, Bayarmagnai Gombodorj, Uuganbayar Zunderiya}
}

\makeatletter
\def\@maketitle{%
  \normalfont\normalsize
  \@settitle
  \@setauthors
  \vskip0.5\baselineskip
  \@setaddresses
  \global\let\addresses\@empty
  \medskip
  \@setabstract
  \medskip
  \begingroup\small
    \noindent\@setkeywords\par
    \medskip\noindent\@setsubjclass\par
  \endgroup
  \bigskip
}
\makeatother

\begin{document}

\begin{abstract}
We study products of multiplication by~$x$ with Riemann--Liouville
fractional integrals and derivatives of any real order. For the integral
$I^\alpha$ of order $\alpha$, the commutator with $x$ gives the relation
$uv-vu=hv^s$ with $s=1+1/\alpha$. The coefficients of
the normal ordering for this relation are the generalized Stirling numbers
$S_{s;h}(n,k)$. We show that these operators give a faithful
representation of the algebra in which the real powers of $v$ are
defined. We expand $(XI^\alpha)^n$ and $(I^\alpha X)^n$ on locally
integrable functions for every $\alpha>0$, and $(XD^\beta)^n$ and
$(D^\beta X)^n$ on the range of $I^{n\beta}$ for every $\beta>0$. The
kernel of $(XI^\alpha)^n$ is an expectation over the Dirichlet
distribution. This gives a probabilistic formula for $S_{s;1}(n,k)$,
$s>1$. By de Finetti's theorem, the formula describes an urn with $n$
colours: fractional integrals correspond to P\'olya urns, and derivatives
of integer order correspond to drawing without replacement. The urn
probabilities are the terms of a product formula that follows from the
recurrence. The cases $s=1,0,2$ give the Stirling numbers of both kinds
and the Bessel numbers. The same method gives the kernels of all words in
$x$ and fractional integrals.
\end{abstract}

\maketitle

\section{Introduction}

Let $X$ denote multiplication by the variable $x$ and $D=d/dx$. The
classical identity $(XD)^n=\sum_{k=1}^nS(n,k)X^kD^k$, with the Stirling
numbers $S(n,k)$ of the second kind, is the basic example of a
normal-ordering formula; see Katriel~\cite{Katriel} and the
monograph~\cite{MansourSchork}. Mansour, Schork, and
Shattuck~\cite{MSS2011} introduced generalized Stirling numbers
$S_{s;h}(n,k)$ through the expansion
\begin{equation}\label{eq:normal-intro}
 (vu)^n=\sum_{k=1}^n S_{s;h}(n,k)v^{s(n-k)+k}u^k,
 \qquad uv-vu=hv^s.
\end{equation}
They did this first for $s\in\Nzero$ and then, by a recurrence, for every
real $s$. Realizations by differential operators with real exponents
were studied in~\cite[Section~4]{MSS2012} and by Maier~\cite{Maier}. For
the Volterra operator $(Vf)(x)=\int_0^xf(t)\,dt$, Benmoussa
\cite[Theorem~1]{BenmoussaJIS} expanded $(XV)^n$ in the operators
$X^kV^{2n-k}$; the coefficients are those of the Bessel polynomials.
Normal orderings of words in $x$ and integrals of integer order were
studied in~\cite{BenmoussaWords}.

This paper is about Riemann--Liouville operators of any real order. The
Riemann--Liouville integrals satisfy
\begin{equation}\label{eq:comm-intro}
 XI^\alpha-I^\alpha X=\alpha I^{\alpha+1}\qquad(\alpha>0).
\end{equation}
Since $I^{\alpha+1}=I^{\alpha(1+1/\alpha)}$, this is the relation
in~\eqref{eq:normal-intro} with $u=X$, $v=I^\alpha$, $h=\alpha$ and
$s=1+1/\alpha$. If $1/\alpha$ is not an integer, then $s$ is not an
integer, and the single relation $uv-vu=hv^s$ does not in general
determine the normal ordering. In this case one needs the commutators of $u$ with all
powers $v^a$, $a\in\Z+\Z s$ (Proposition~\ref{prop:compatibility}). For
Riemann--Liouville operators these powers are given by the semigroup
$\{I^r\}$, $v^a=I^{\alpha a}$, and fractional derivatives are the powers
of negative order. Our results are of two kinds.

\emph{Fractional operators.}
\begin{enumerate}[\upshape(i)]
\item For every $s\ne1$, the assignment $v^a\mapsto I^{a/(s-1)}$,
$u\mapsto h(s-1)X$ is a faithful representation of the algebra $\As$ in
which the powers $v^a$ are defined. It acts on smooth functions that are
flat at the origin, and integrals of negative order are
Riemann--Liouville derivatives (Theorem~\ref{thm:realization}). Under the
Laplace transform it becomes the well-known differential realization
(Remark~\ref{rem:laplace}).
\item For every $\alpha>0$ and $n\ge1$, on locally integrable functions,
\begin{equation}\label{eq:main-intro}
 (XI^\alpha)^n=\sum_{k=1}^n S_{1+1/\alpha;-\alpha}(n,k)\,
                  X^kI^{(\alpha+1)n-k},
\end{equation}
and $(I^\alpha X)^n$ has the expansion in the opposite order with
coefficients $S_{1+1/\alpha;\alpha}(n,k)$. For every $\beta>0$, the
corresponding expansions of $(XD^\beta)^n$ and $(D^\beta X)^n$, with
coefficients $S_{1-1/\beta;\pm\beta}(n,k)$, hold on the range of
$I^{n\beta}$ (Theorem~\ref{thm:L1}). For derivatives, a condition at the
origin is needed (Remark~\ref{rem:domain}). The coefficients are
explicit polynomials in $\alpha$ (Proposition~\ref{prop:explicit}).
Products $I^\gamma XI^\delta$ lead to the full three-parameter family of
Hsu and Shiue~\cite{HsuShiue} (Proposition~\ref{prop:three}). The cases
$\alpha=1$, $\beta=1$ and $\beta=\frac12$ give Benmoussa's expansion,
Scherk's identity and the Bessel numbers of the second kind
(Corollary~\ref{cor:special}).
\end{enumerate}

\emph{A probabilistic formula.}
\begin{enumerate}[\upshape(i)]\setcounter{enumi}{2}
\item For $\alpha>0$, the kernel of $(XI^\alpha)^n$ is
$x^n(x-t)^{n\alpha-1}P_{n,\alpha}(1-t/x)/\Gamma(n\alpha)$ with
$P_{n,\alpha}(y)=\mathbb E\prod_{j=1}^{n-1}(1-yB_j)$. Here the $B_j$
are the partial sums of a symmetric Dirichlet vector with parameter
$\alpha$ (Theorem~\ref{thm:kernel}). As a consequence, for every real
$s>1$,
\begin{equation}\label{eq:prob-intro}
 S_{s;1}(n,n-m)=\mathbb E\,e_m(B_1,\dots,B_{n-1})
     \prod_{i=0}^{m-1}\bigl(n+i(s-1)\bigr),
\end{equation}
where $e_m$ is the elementary symmetric polynomial and the Dirichlet
parameter is $1/(s-1)$ (Corollary~\ref{cor:dirichlet}). The reversal
symmetry of the Dirichlet distribution corresponds to switching between
the two normal orders.
\item By de Finetti's theorem, for $s>1$ (fractional integrals),
\eqref{eq:prob-intro} is a statement about a P\'olya urn with $n$
colours, in which each drawn colour becomes more likely to be drawn
again. The urn model also works for $s=1$, where the draws are
independent, and for $s=1-1/K$ with $K\in\N$ (derivatives of order $K$),
where the balls are drawn without replacement from an urn with $K$ balls
of each colour (Proposition~\ref{prop:urn}). The urn
probabilities are the terms of a product formula for $S_{s;1}(n,n-m)$.
This formula follows from the recurrence and holds for every $s$. At
$s=1$, $s=0$ and $s=2$ it gives the classical formulas for the Stirling
numbers of the first and second kinds and for the Bessel numbers
(Corollary~\ref{cor:urn-special}). For $s\ge1$ the same numbers also
describe the number of blocks of the Ewens--Pitman partition, which is a
different urn model (Remark~\ref{rem:pitman}).
\item The same method gives the kernels of all words
$X^{\lambda_n}I^{\alpha_n}\cdots X^{\lambda_1}I^{\alpha_1}$ with
$\alpha_i>0$ and $\lambda_i\in\Nzero$ (Theorem~\ref{thm:words}). They are
expectations over Dirichlet distributions that are not symmetric, the
coefficients of the normal ordering are explicit with alternating signs,
and the urn model extends to these words. For $(X^\lambda I^\alpha)^n$
this extends a recurrence of Benmoussa~\cite{BenmoussaWords} from integer
to real orders (Corollary~\ref{cor:power-words}).
\end{enumerate}

Butzer, Kilbas, and Trujillo~\cite{BKT} use fractional calculus to extend
the Stirling numbers of the second kind to non-integer values of the
first index; this is a different kind of extension. Up to a power of
$x$, $(XI^\alpha)^n$ is a multiple Erd\'elyi--Kober operator in the sense
of Kiryakova~\cite{Kiryakova}. Such operators have Meijer $G$-function
kernels in general, but by~\eqref{eq:main-intro} the kernel is
elementary here. Under the Laplace transform, the identities of~(ii)
correspond to identities for differential operators that are contained
in~\cite{Maier} (Remark~\ref{rem:laplace}). We do not know of earlier
normal-ordering formulas for Riemann--Liouville operators of non-integer
order, or of a description of the kernels of $(XI^\alpha)^n$ and of other
words in $X$ and fractional integrals by Dirichlet distributions or by
P\'olya urns.

Throughout, algebras are unital associative $\C$-algebras,
homomorphisms preserve the identity, $h\in\C\setminus\{0\}$,
and $[a,b]=ab-ba$. We write $\Nzero=\{0,1,2,\ldots\}$,
$\N=\Nzero\setminus\{0\}$, and $(a)_m=a(a+1)\cdots(a+m-1)$. All sums
are finite, and empty products equal~$1$.

\section{Generalized Stirling numbers and compatible powers}\label{sec:algebra}

For $s\in\R$, define $S_{s;h}(0,0)=1$, $S_{s;h}(n,0)=0$ for
$n\ge1$, $S_{s;h}(n,k)=0$ for $k<0$ or $k>n$, and
\begin{equation}\label{eq:recurrence}
 S_{s;h}(n+1,k)=S_{s;h}(n,k-1)
       +h\bigl(k+s(n-k)\bigr)S_{s;h}(n,k)\qquad(n\ge0,\ k\ge1),
\end{equation}
as in~\cite[Definition~3.1]{MSS2011}. Induction gives
\begin{equation}\label{eq:scaling}
 S_{s;h}(n,k)=h^{n-k}S_{s;1}(n,k),
 \qquad
 S_{s;-h}(n,k)=(-1)^{n-k}S_{s;h}(n,k).
\end{equation}
For example, $S_{s;h}(n,n)=1$, $S_{s;h}(2,1)=h$, $S_{s;h}(3,2)=3h$,
and $S_{s;h}(3,1)=h^2(s+1)$. The numbers $S_{s;h}(n,k)$ are connection
coefficients, as the next lemma shows.

\begin{lemma}\label{lem:connection}
For $s\in\R$ and $n\in\Nzero$, as polynomials in $z$,
\begin{equation}\label{eq:connection}
 \prod_{i=0}^{n-1}(z+hsi)
   =\sum_{k=0}^n S_{s;h}(n,k)\prod_{j=0}^{k-1}\bigl(z-h(1-s)j\bigr).
\end{equation}
\end{lemma}
\begin{proof}
The case $n=0$ is clear. Multiply~\eqref{eq:connection} by
$z+hsn=\bigl(z-h(1-s)k\bigr)+h\bigl(k+s(n-k)\bigr)$
and compare with~\eqref{eq:recurrence}.
\end{proof}

Hsu and Shiue~\cite{HsuShiue} defined numbers $S(n,k;a,b,r)$ by
\begin{equation}\label{eq:HS}
 (z\,|\,a)_n=\sum_{k=0}^n S(n,k;a,b,r)\,(z-r\,|\,b)_k,
 \qquad (z\,|\,a)_n=\prod_{j=0}^{n-1}(z-ja).
\end{equation}
Thus \eqref{eq:connection} says that
$S_{s;h}(n,k)=S(n,k;-hs,h(1-s),0)$; this was observed
in~\cite[Theorem~2]{MSS2012}.

\subsection{\texorpdfstring{The algebra $\As$}{The algebra A(s,h)}}
Let $\Gs=\Z+\Z s\subseteq\R$ and $\Rs=\C[\Gs]$, where we view $\Gs$ as an
additive subgroup of $\R$. Thus $\Rs$ has basis $\{v^a:a\in\Gs\}$ and
\begin{equation}\label{eq:powers}
 v^av^b=v^{a+b},\qquad v^0=1.
\end{equation}
If $s=p/q$ in lowest terms, with $q>0$, then $\Gs=q^{-1}\Z$ and
$\Rs=\C[t,t^{-1}]$ with $t=v^{1/q}$. If $s$ is irrational, then
$\Rs=\C[x^{\pm1},y^{\pm1}]$ with $x=v$ and $y=v^s$. The linear map
$\delta_{s,h}(v^a)=ha\,v^{a+s-1}$ is a derivation of $\Rs$, and we
consider the differential Ore extension
\begin{equation}\label{eq:Ore}
 \As=\Rs[u;\delta_{s,h}],
 \qquad ur=ru+\delta_{s,h}(r)\quad(r\in\Rs);
\end{equation}
see~\cite{Ore,GoodearlWarfield}. Equivalently, in addition to
\eqref{eq:powers}, its generators satisfy
\begin{equation}\label{eq:compatible}
 [u,v^a]=ha\,v^{a+s-1}\qquad(a\in\Gs).
\end{equation}
By the universal property of Ore extensions, any elements of another
algebra that satisfy \eqref{eq:powers} and \eqref{eq:compatible}
determine a unique homomorphism from $\As$.

\begin{lemma}\label{lem:normal-forms}
Both $\{v^au^m:a\in\Gs,\ m\in\Nzero\}$ and
$\{u^mv^a:a\in\Gs,\ m\in\Nzero\}$ are $\C$-bases of $\As$.
\end{lemma}
\begin{proof}
The first statement is part of the construction of the Ore extension.
With $\delta=\delta_{s,h}$, induction gives
$u^mr=\sum_{j=0}^m\binom mj\delta^j(r)u^{m-j}$ and
$ru^m=\sum_{j=0}^m(-1)^j\binom mju^{m-j}\delta^j(r)$ for $r\in\Rs$.
The second identity shows that the elements $u^mv^a$ span the algebra. If
$\sum_{m=0}^Mu^mr_m=0$ with $r_M\ne0$, the first identity shows that the
coefficient of $u^M$ in the first normal form is $r_M$, a contradiction.
\end{proof}

\begin{theorem}\label{thm:ordering}
Let $s\in\R$ and $n\ge1$. In $\As$,
\begin{align}
 (vu)^n&=\sum_{k=1}^n S_{s;h}(n,k)v^{s(n-k)+k}u^k,
       \label{eq:normal}\\
 (uv)^n&=\sum_{k=1}^n S_{s;-h}(n,k)u^kv^{s(n-k)+k}
        =\sum_{k=1}^n(-1)^{n-k}S_{s;h}(n,k)u^kv^{s(n-k)+k}.
       \label{eq:reverse}
\end{align}
\end{theorem}
\begin{proof}
Set $e_{n,k}=s(n-k)+k$. By \eqref{eq:compatible},
$vu\,v^{e_{n,k}}u^k=v^{e_{n,k}+1}u^{k+1}+he_{n,k}v^{e_{n,k}+s}u^k$.
Since $e_{n,k}+1=e_{n+1,k+1}$ and $e_{n,k}+s=e_{n+1,k}$, the
coefficients satisfy \eqref{eq:recurrence}, and \eqref{eq:normal}
follows by induction. In the opposite algebra, with product
$a\circ b=ba$, the powers satisfy \eqref{eq:powers} and
$u\circ v^a-v^a\circ u=-ha\,v^{a+s-1}$. So the identity map on the
generators gives a homomorphism $A_{s,-h}\to\As^{\mathrm{op}}$, and this
homomorphism is bijective by Lemma~\ref{lem:normal-forms}. Applying \eqref{eq:normal} in
$A_{s,-h}$ gives \eqref{eq:reverse}, and \eqref{eq:scaling} gives its
second form.
\end{proof}

\begin{remark}\label{rem:known}
For the differential operators $u=x^s\,d/dx$ and $v=x$ with complex $s$,
identities equivalent to \eqref{eq:normal} and \eqref{eq:reverse} are
special cases of the expansions of Maier~\cite[Theorem~6.4]{Maier}. For
positive integer $s$, the expansion \eqref{eq:reverse} was obtained by
Burde~\cite[Proposition~5.3]{Burde} and in Benmoussa's
preprint~\cite[Theorem~4.1]{BenmoussaPreprint}. The simplest case,
$(uv)^2=u(uv-hv^s)v=u^2v^2-huv^{s+1}$, shows that the coefficients in
\cite[Conjecture~2]{BenmoussaJIS} need $-h$ in place of $h$.
\end{remark}

\subsection{When are compatible powers needed?}
For integer $s$ the relation $[u,v]=hv^s$ is enough.

\begin{corollary}\label{cor:integer}
Let $u,v$ belong to an algebra and satisfy $[u,v]=hv^s$.
If $s\in\Nzero$, the identities \eqref{eq:normal} and
\eqref{eq:reverse} hold without any invertibility assumption.
They also hold for negative integer $s$ if $v$ is invertible.
\end{corollary}
\begin{proof}
For $a\in\N$, the Leibniz rule gives
$[u,v^a]=\sum_{j=0}^{a-1}v^j[u,v]v^{a-1-j}=ha\,v^{a+s-1}$, and, if $v$
is invertible, $[u,v^{-1}]=-v^{-1}[u,v]v^{-1}=-hv^{s-2}$. The proof of
Theorem~\ref{thm:ordering} uses only these commutators, because all
exponents $e_{n,k}$ are integers, and they are positive if $s\in\Nzero$.
\end{proof}

For non-integer $s$ this is no longer true.

\begin{proposition}\label{prop:compatibility}
Let an algebra $B$ contain $\Rs$ as a unital subalgebra, and
let $u\in B$ satisfy $[u,v]=hv^s$.
\begin{enumerate}[\upshape(1)]
\item If $s=p/q$ in lowest terms, with $q>0$, then
\eqref{eq:compatible} holds for every $a\in\Gs$ if and only if
$[u,v^{1/q}]$ commutes with $v^{1/q}$. For $s=1/2$, for example, this
condition can fail, and then \eqref{eq:reverse} can fail too.
\item If $s$ is irrational, then \eqref{eq:compatible} holds
for every $a\in\Gs$ if and only if
\begin{equation}\label{eq:extra}
 [u,v^s]=hs\,v^{2s-1}.
\end{equation}
The assumption that all $[u,v^a]$, $a\in\Gs$, commute with $\Rs$ is not
enough, and in that case \eqref{eq:reverse} can fail.
\end{enumerate}
\end{proposition}
\begin{proof}
(1) Put $t=v^{1/q}$ and $c=[u,t]$. If $c$ commutes with $t$, then
$ht^p=[u,t^q]=qt^{q-1}c$, so $c=\frac hq\,t^{p-q+1}$, and the Leibniz
rule gives $[u,t^j]=(hj/q)t^{j+p-q}$ for all $j\in\Z$; this is
\eqref{eq:compatible} for $a=j/q$. Conversely, \eqref{eq:compatible}
with $a=1/q$ gives this value of $c$. For the example, let $s=1/2$, so
that $\Rs=\C[t,t^{-1}]$, and embed $\Rs$ in $\End_\C(\Rs)$ by
multiplication operators. Let $(Qf)(t)=f(-t)$, so that $Qt=-tQ$, and put
$u=\frac h2\frac d{dt}+\kappa\,Q\circ\frac1{2t}$ with $\kappa\ne0$. Then
$c=[u,t]=\frac h2+\kappa Q$ does not commute with $t$, while
$[u,v]=ct+tc=ht=hv^s$. For $n=3$, the left side of~\eqref{eq:reverse}
minus the right side maps $1$ to $\frac12h\kappa(4h-\kappa)t^3$ and $t$
to $-\frac12h\kappa(5h+\kappa)t^4$, and these are not both zero.
(2) Since $\Rs$ is generated by $v^{\pm1}$ and $v^{\pm s}$, the two
derivations $[u,\cdot\,]$ and $\delta_{s,h}$ agree on $\Rs$ once they
agree on $v$ and $v^s$. For the last statement, embed
$\Rs=\C[x^{\pm1},y^{\pm1}]$ in $\End_\C(\Rs)$ by multiplication
operators and take $u=hy\,\partial/\partial x$. Then
$[u,v^{m+ns}]=hm\,v^{m+ns+s-1}$ for $m,n\in\Z$: these commutators lie
in $\Rs$, $[u,v]=hv^s$, but \eqref{eq:compatible} fails for $n\ne0$.
For $n=3$, the left side of \eqref{eq:reverse} maps $1$ to $h^3y^3$,
while the right side,
$u^3v^3-3hu^2v^{s+2}+h^2(s+1)uv^{2s+1}$, maps $1$ to
$6h^3y^3-6h^3y^3+(s+1)h^3y^3=(s+1)h^3y^3$.
\end{proof}

In the next sections, fractional calculus gives the compatible powers.

\section{Riemann--Liouville operators}\label{sec:RL}

Let $\Lloc$ be the space of measurable functions
$f\colon(0,\infty)\to\C$ that are integrable on $(0,T)$ for every
$T>0$; we identify functions that are equal almost everywhere. Let $X$
denote multiplication by $x$. For $r>0$, the Riemann--Liouville
integral of order $r$ is
\begin{equation}\label{eq:RL}
 (I^rf)(x)=\frac1{\Gamma(r)}\int_0^x(x-t)^{r-1}f(t)\,dt,
\end{equation}
and $I^0=\Id$; see~\cite{KST,SKM}. Both $X$ and $I^r$ map $\Lloc$
into itself. For $\beta>0$ and $m=\lceil\beta\rceil$, the
Riemann--Liouville derivative is $D^\beta f=(d/dx)^mI^{m-\beta}f$,
defined when $I^{m-\beta}f$ has a locally absolutely continuous
derivative of order $m-1$ on $[0,\infty)$. Explicitly, the domain
$\Dom(D^\beta)$ consists of all $f\in\Lloc$ such that
\begin{equation}\label{eq:domain}
 I^{m-\beta}f=p+I^m\psi
\end{equation}
for a polynomial $p$ of degree less than $m$ and some $\psi\in\Lloc$,
and then $D^\beta f=\psi$. For $\beta\in\N$ this is the ordinary
derivative. We also write $D^{-r}=I^r$ for $r\ge0$.

\begin{lemma}\label{lem:L1}
Let $r,q\ge0$ and $\gamma\ge\beta>0$. On $\Lloc$,
\begin{enumerate}[\upshape(a)]
\item $I^rI^q=I^{r+q}$, and $I^r$ is injective;
\item $XI^r-I^rX=rI^{r+1}$;
\item $I^\gamma\varphi\in\Dom(D^\beta)$ and
$D^\beta I^\gamma\varphi=I^{\gamma-\beta}\varphi$ for every
$\varphi\in\Lloc$.
\end{enumerate}
\end{lemma}
\begin{proof}
(a) The semigroup law follows from Fubini's theorem and the beta
integral \cite[Section~2.1]{KST}. If $I^r\varphi=0$, choose an integer
$m\ge r$; then $I^m\varphi=I^{m-r}I^r\varphi=0$, and differentiating
$m$ times gives $\varphi=0$.
(b) For $r>0$, subtracting the integrals gives
\[
 (XI^rf-I^rXf)(x)=\frac1{\Gamma(r)}\int_0^x(x-t)^rf(t)\,dt
                 =r\,(I^{r+1}f)(x),
\]
and for $r=0$ both sides are zero.
(c) With $m=\lceil\beta\rceil$, part~(a) gives
$I^{m-\beta}I^\gamma\varphi=I^m\psi$, where
$\psi=I^{\gamma-\beta}\varphi\in\Lloc$. This is~\eqref{eq:domain}
with $p=0$.
\end{proof}

On the space
\[
 \E=\{f\in C^\infty([0,\infty);\C):
          f^{(j)}(0)=0\ \text{for all }j\in\Nzero\}
\]
of smooth functions that are flat at the origin, operators of all real
orders can be composed, and their orders add. This space contains
$C_c^\infty((0,\infty))$ and is contained in $\Lloc$. Let $D=d/dx$.

\begin{lemma}\label{lem:flat}
The operators $X$, $D$, and $I^r$ $(r>0)$ map $\E$ into itself, and
$DI^r=I^rD$ on $\E$. For $r\le0$ and $f\in\E$ define
$I^rf=D^mI^{r+m}f$, where $m\in\Nzero$ and $m\ge-r$. This does not
depend on $m$, and on $\E$
\begin{equation}\label{eq:group}
 I^0=\Id,\qquad I^rI^q=I^{r+q},\qquad
 XI^r-I^rX=rI^{r+1}\qquad(r,q\in\R).
\end{equation}
In particular, every $I^r$ is a bijection of $\E$, and $I^{-\beta}$ is
the Riemann--Liouville derivative $D^\beta$ on $\E$ for $\beta>0$.
\end{lemma}
\begin{proof}
Let $r>0$ and $f\in\E$. For continuous $g$, the function
$I^rg(x)=x^r\Gamma(r)^{-1}\int_0^1(1-\tau)^{r-1}g(x\tau)\,d\tau$
is continuous on $[0,\infty)$ and vanishes at $0$. Since $f(0)=0$,
we have $f=I^1f'$, hence $I^rf=I^1(I^rf')$ by
Lemma~\ref{lem:L1}(a). Thus $I^rf\in C^1([0,\infty))$ and
$DI^rf=I^rDf$. As $f'\in\E$, induction gives
$I^rf\in C^\infty([0,\infty))$ and $D^jI^rf=I^rD^jf$ for all $j$,
and these functions vanish at $0$. Hence $I^r\E\subseteq\E$ and
$DI^r=I^rD$, and $X$ and $D$ clearly map this space into itself.

Since $DI^1=\Id$ on $\E$, we have
$D^{m+1}I^{r+m+1}=D^mI^{r+m}$ whenever $r+m\ge0$. Hence $I^r$ does not
depend on $m$, and $I^r=D^mI^{r+m}$ for all $r\in\R$ and all
integers $m\ge\max\{0,-r\}$. As $D$ commutes with the integrals of
positive order,
\[
 I^rI^q=D^mI^{r+m}D^{m'}I^{q+m'}
       =D^{m+m'}I^{r+q+m+m'}=I^{r+q}
\]
for large $m,m'$, so $I^r$ is bijective with inverse $I^{-r}$. For the
commutator, let $m\ge\max\{1,-r\}$. By Lemma~\ref{lem:L1}(b)
and $[X,D^m]=-mD^{m-1}$,
\[
 [X,D^mI^{r+m}]=-mD^{m-1}I^{r+m}+(r+m)D^mI^{r+m+1}
               =-mI^{r+1}+(r+m)I^{r+1}=rI^{r+1}.
\]
Finally, let $\beta>0$, $m=\lceil\beta\rceil$, and $f\in\E$. Then
$g=I^{m-\beta}f\in\E$ satisfies $g=I^mD^mg$, which
is~\eqref{eq:domain} with $p=0$; hence
$D^\beta f=D^mI^{m-\beta}f=I^{-\beta}f$.
\end{proof}

\section{The fractional realization and normal ordering}\label{sec:realization}

\begin{theorem}\label{thm:realization}
Let $s\in\R\setminus\{1\}$, $h\in\C\setminus\{0\}$, and
$\alpha=1/(s-1)$. There is a unique homomorphism
$\rho\colon\As\to\End_\C(\E)$ such that
\[
 \rho(v^a)=I^{\alpha a}\quad(a\in\Gs),\qquad
 \rho(u)=\frac h\alpha\,X,
\]
and $\rho$ is injective. Here $\rho(v)=I^\alpha$ is a fractional
integral if $s>1$ and the fractional derivative $D^{-\alpha}$ if $s<1$.
\end{theorem}
\begin{proof}
By Lemma~\ref{lem:flat}, the operators $I^{\alpha a}$
satisfy~\eqref{eq:powers}, and
\[
 \Bigl[\frac h\alpha X,I^{\alpha a}\Bigr]
   =ha\,I^{\alpha a+1}=ha\,I^{\alpha(a+s-1)},
\]
because $\alpha(s-1)=1$. This is~\eqref{eq:compatible}, so $\rho$
exists and is unique by the universal property of $\As$.

Let $P=\sum_{(a,k)\in F}c_{a,k}v^au^k$, with
$F\subseteq\Gs\times\Nzero$ finite, and suppose that $\rho(P)=0$.
With $c'_{a,k}=c_{a,k}(h/\alpha)^k$ this means that
$\sum_{F}c'_{a,k}I^{\alpha a}X^k=0$ on $\E$. For $\lambda>0$ the
dilation $(S_\lambda f)(x)=f(\lambda x)$ is a bijection of $\E$, and
\[
 XS_\lambda=\lambda^{-1}S_\lambda X,\qquad
 I^rS_\lambda=\lambda^{-r}S_\lambda I^r\quad(r\in\R);
\]
the second relation follows from~\eqref{eq:RL} for $r>0$ and then,
by taking inverses, for $r<0$. Hence, for every $f\in\E$,
\[
 0=\rho(P)S_\lambda f
  =S_\lambda\sum_{(a,k)\in F}\lambda^{-\alpha a-k}
      c'_{a,k}I^{\alpha a}X^kf\qquad(\lambda>0).
\]
Since $S_\lambda$ is injective and the functions
$\lambda\mapsto\lambda^{-d}$ with distinct real $d$ are linearly
independent, we obtain, for every $d\in\R$,
\[
 \sum_{k}b_kI^{d-k}X^k=0\ \text{ on }\E,
 \qquad\text{where } b_k=c'_{(d-k)/\alpha,\,k}
\]
and $b_k=0$ if $((d-k)/\alpha,k)\notin F$. Applying $I^{-d}$ gives
$\sum_kb_kD^kX^k=0$ on $\E$. Suppose that some $b_k$ is nonzero, and
let $K$ be the largest such $k$. Since
$D^kX^k=\sum_{j=0}^k\binom kj\frac{k!}{j!}X^jD^j$, the operator
$\sum_kb_kD^kX^k$ has the form $\sum_{j=0}^Ka_jX^jD^j$ with
$a_K=b_K\ne0$. Applied to a function $f\in C_c^\infty((0,\infty))$
with $f(x)=(x-1)^K$ near $x=1$, it gives a function with the value
$a_KK!\ne0$ at $x=1$, a contradiction. Hence all $b_k$ are zero, for
every $d$. Since $\alpha\ne0$, each pair $(a,k)\in F$ occurs for
$d=\alpha a+k$, and therefore $P=0$.
\end{proof}

Thus the relations of $\As$ are exactly the relations between $X$ and a
group of fractional integrals and derivatives. For $s=1$ this
construction does not apply, since $[X,I^r]$ has order $r+1$.

\begin{corollary}\label{cor:flat}
Let $\alpha\in\R\setminus\{0\}$, $s=1+1/\alpha$, and $n\ge1$.
On $\E$,
\begin{align}
 (I^\alpha X)^n&=\sum_{k=1}^n S_{s;\alpha}(n,k)\,
                  I^{(\alpha+1)n-k}X^k,\label{eq:IX}\\
 (XI^\alpha)^n&=\sum_{k=1}^n S_{s;-\alpha}(n,k)\,
                  X^kI^{(\alpha+1)n-k}.\label{eq:XI}
\end{align}
The operators $X^kI^r$ $(k\in\Nzero,\ r\in\Z+\Z\alpha)$ are linearly
independent on $\E$, and so are the operators $I^rX^k$. In particular,
the coefficients in~\eqref{eq:IX} and~\eqref{eq:XI} are unique.
\end{corollary}
\begin{proof}
Take $h=\alpha$ in Theorem~\ref{thm:realization}. Then
$\rho(u)=X$, $\rho(v)=I^\alpha$, and
$\rho(v^{s(n-k)+k})=I^{(\alpha+1)n-k}$, because $\alpha s=\alpha+1$.
Applying $\rho$ to \eqref{eq:normal} and~\eqref{eq:reverse} gives
\eqref{eq:IX} and~\eqref{eq:XI}. The independence follows from
Lemma~\ref{lem:normal-forms} and the injectivity of $\rho$.
\end{proof}

To pass from $\E$ to integrable functions, we use only operators of
nonnegative order. Fix $\alpha\in\R\setminus\{0\}$, let
$s=1+1/\alpha$ and $G=\Z+\Z\alpha=\alpha\Gs$, and write
$J_r=v^{r/\alpha}$ for $r\in G$ in the algebra $A=A_{s,\alpha}$.
Then \eqref{eq:powers} and~\eqref{eq:compatible} become
\begin{equation}\label{eq:additive}
 J_0=1,\qquad J_rJ_q=J_{r+q},\qquad [u,J_r]=rJ_{r+1}
 \qquad(r,q\in G),
\end{equation}
and $v=J_\alpha$, $v^{s(n-k)+k}=J_{(\alpha+1)n-k}$. For $c\ge0$ let
$A_{\ge c}$ be the span of
$\{J_ru^m:r\in G,\ r\ge c,\ m\in\Nzero\}$.

\begin{lemma}\label{lem:positive}
\begin{enumerate}[\upshape(a)]
\item $A_{\ge0}$ is a subalgebra of $A$. Moreover,
$uA_{\ge c}\subseteq A_{\ge c}$ and
$J_rA_{\ge c}\subseteq A_{\ge c+r}$ for $c\ge0$ and $r\in G$ with
$c+r\ge0$.
\item There is a unique homomorphism
$\sigma\colon A_{\ge0}\to\End_\C(\Lloc)$ such that $\sigma(J_r)=I^r$
for $r\in G\cap[0,\infty)$ and $\sigma(u)=X$.
\item Let $\gamma\in G$, $\gamma>0$, $w\in A_{\ge\gamma}$, and
$\varphi\in\Lloc$. Then $\sigma(w)\varphi\in\Dom(D^\gamma)$ and
$D^\gamma\sigma(w)\varphi=\sigma(J_{-\gamma}w)\varphi$.
\end{enumerate}
\end{lemma}
\begin{proof}
(a) By~\eqref{eq:additive}, $u\,J_ru^m=J_ru^{m+1}+rJ_{r+1}u^m$ and
$J_r\,J_qu^m=J_{r+q}u^m$. This gives the two inclusions, and they
imply that $A_{\ge0}$ is closed under multiplication.
(b) The span $R_+$ of $\{J_r:r\in G,\ r\ge0\}$ is a subalgebra of
$\Rs$ with $\delta_{s,\alpha}(R_+)\subseteq R_+$, and
$A_{\ge0}=\bigoplus_mR_+u^m$ is the Ore extension of $R_+$ by the
restricted derivation. By Lemma~\ref{lem:L1}(a), $J_r\mapsto I^r$
defines a homomorphism $R_+\to\End_\C(\Lloc)$, and by
Lemma~\ref{lem:L1}(b), $[X,I^r]=rI^{r+1}$ is the image of
$\delta_{s,\alpha}(J_r)=rJ_{r+1}$. The universal property of Ore
extensions~\cite[Chapter~2]{GoodearlWarfield} gives~$\sigma$.
(c) By linearity we may assume that $w=J_ru^m$ with $r\ge\gamma$.
Then $\sigma(w)\varphi=I^r(X^m\varphi)$, and Lemma~\ref{lem:L1}(c)
gives $D^\gamma I^r(X^m\varphi)=I^{r-\gamma}X^m\varphi
=\sigma(J_{-\gamma}w)\varphi$.
\end{proof}

\begin{theorem}\label{thm:L1}
Let $n\ge1$.
\begin{enumerate}[\upshape(a)]
\item Let $\alpha>0$ and $s=1+1/\alpha$. Then \eqref{eq:IX}
and~\eqref{eq:XI} hold on $\Lloc$.
\item Let $\beta>0$ and $f=I^{n\beta}\varphi$ with $\varphi\in\Lloc$.
Then every Riemann--Liouville derivative below is applied to a function
in its domain. Moreover, $D^{(\beta-1)n+k}f=I^{n-k}\varphi$ for
$1\le k\le n$, and
\begin{align}
 (D^\beta X)^nf&=\sum_{k=1}^n S_{1-1/\beta;-\beta}(n,k)\,
                  D^{(\beta-1)n+k}X^kf,\label{eq:DX}\\
 (XD^\beta)^nf&=\sum_{k=1}^n S_{1-1/\beta;\beta}(n,k)\,
                  X^kD^{(\beta-1)n+k}f.\label{eq:XD}
\end{align}
\end{enumerate}
\end{theorem}
\begin{proof}
(a) Here $u$, $v=J_\alpha$, and all terms of \eqref{eq:normal}
and~\eqref{eq:reverse} lie in $A_{\ge0}$, because
$(\alpha+1)n-k\ge0$. Applying $\sigma$ to Theorem~\ref{thm:ordering}
for $A_{s,\alpha}$ gives the result.

(b) We use Lemma~\ref{lem:positive} with $\alpha=-\beta$. Then
$s=1-1/\beta$, $h=-\beta$, $v=J_{-\beta}$, and
$f=\sigma(J_{n\beta})\varphi$. We
claim that, for $0\le j\le n$,
\[
 (D^\beta X)^jf=\sigma(w_j)\varphi,\qquad
 w_j=(vu)^jJ_{n\beta}\in A_{\ge(n-j)\beta}.
\]
This is clear for $j=0$. If it holds for some $j<n$, then
$uw_j\in A_{\ge(n-j)\beta}$ and
$X\sigma(w_j)\varphi=\sigma(uw_j)\varphi$. Since $(n-j)\beta\ge\beta$,
Lemma~\ref{lem:positive}(c) shows that this function lies in
$\Dom(D^\beta)$ and that
$D^\beta\sigma(uw_j)\varphi=\sigma(J_{-\beta}uw_j)\varphi
=\sigma(w_{j+1})\varphi$, and $w_{j+1}\in A_{\ge(n-j-1)\beta}$ by
Lemma~\ref{lem:positive}(a). In the same way,
$(XD^\beta)^jf=\sigma\bigl((uv)^jJ_{n\beta}\bigr)\varphi$.

Put $\gamma_k=(\beta-1)n+k$, so that $v^{s(n-k)+k}=J_{-\gamma_k}$ and
$n\beta-\gamma_k=n-k$. Theorem~\ref{thm:ordering} for $A_{s,-\beta}$
gives
\[
 (vu)^nJ_{n\beta}=\sum_{k=1}^nS_{s;-\beta}(n,k)\,
                    J_{-\gamma_k}u^kJ_{n\beta},
 \qquad
 (uv)^nJ_{n\beta}=\sum_{k=1}^nS_{s;\beta}(n,k)\,u^kJ_{n-k}.
\]
Here $u^kJ_{n\beta}\in A_{\ge n\beta}$ and $\gamma_k\le n\beta$. Hence
$\sigma(J_{-\gamma_k}u^kJ_{n\beta})\varphi=D^{\gamma_k}X^kf$ and
$\sigma(J_{n-k})\varphi=D^{\gamma_k}f$, by
Lemma~\ref{lem:positive}(c) if $\gamma_k>0$ and by the convention
$D^{\gamma_k}=I^{-\gamma_k}$ if $\gamma_k\le0$. Applying $\sigma$ to
the two identities gives \eqref{eq:DX} and~\eqref{eq:XD}.
\end{proof}

The same statements hold on a finite interval $(0,T)$. The first cases
are
\begin{align*}
 (XI^\alpha)^2&=X^2I^{2\alpha}-\alpha XI^{2\alpha+1},&
 (XI^\alpha)^3&=X^3I^{3\alpha}-3\alpha X^2I^{3\alpha+1}
                +\alpha(2\alpha+1)XI^{3\alpha+2},\\
 (XD^\beta)^2&=X^2D^{2\beta}+\beta XD^{2\beta-1},&
 (XD^\beta)^3&=X^3D^{3\beta}+3\beta X^2D^{3\beta-1}
                +\beta(2\beta-1)XD^{3\beta-2}.
\end{align*}

\begin{remark}\label{rem:domain}
For $\beta\in\N$ both sides of \eqref{eq:DX} and~\eqref{eq:XD} are
differential operators with polynomial coefficients; they agree on
$C_c^\infty((0,\infty))\subseteq\E$ by Corollary~\ref{cor:flat}, hence
on $C^{n\beta}((0,\infty))$. For non-integer $\beta$, a condition at the
origin is needed. Let $1<\beta<2$ and $f(x)=x^{\beta-2}$. Then
$I^{2-\beta}f$ is constant, so $D^\beta f=0$ and $(XD^\beta)^nf=0$ for
all $n\ge1$. However, $f\notin\Dom(D^{2\beta})$, because
$I^{m-2\beta}f$ is a nonzero multiple of $x^{m-\beta-2}$ for
$m=\lceil2\beta\rceil$, and this function is not of the
form~\eqref{eq:domain}. So the right-hand side of~\eqref{eq:XD} is not
defined for $n=2$ in the sense of Section~\ref{sec:RL}. If its
derivatives are computed pointwise on $(0,\infty)$, that is, by
$D^\gamma x^\mu=\Gamma(\mu+1)x^{\mu-\gamma}/\Gamma(\mu+1-\gamma)$ for
$\mu>-1$, it equals
\[
 x^2D^{2\beta}f+\beta xD^{2\beta-1}f
   =\Gamma(\beta-1)\Bigl(\frac1{\Gamma(-\beta-1)}
        +\frac\beta{\Gamma(-\beta)}\Bigr)x^{-\beta}
   =-\frac{\Gamma(\beta-1)}{\Gamma(-\beta)}\,x^{-\beta}\ne0.
\]
Thus $D^\beta D^\beta f\ne D^{2\beta}f$, even pointwise. On $\Lloc$ the
Riemann--Liouville derivatives do not form a semigroup, and the
operators $D^{\beta a}$ cannot be used as the powers $v^a$ of
$v=D^\beta$ in~\eqref{eq:powers}. On the other hand, the range of
$I^{n\beta}$ is a sufficient domain but not the largest one. For
example, for $n=2$ and $f(x)=x^{2\beta-1}$, which is not in the range of
$I^{2\beta}$, both sides of~\eqref{eq:XD} are defined and equal
$\beta\Gamma(2\beta)x$.
\end{remark}

\begin{remark}\label{rem:integral-s}
If $\alpha=1/m$ with $m\in\N$, then $s=m+1$ and $J=I^{1/m}$ satisfies
$[X,J]=\frac1mJ^{m+1}$, so that \eqref{eq:IX} and~\eqref{eq:XI} follow
from this single relation by Corollary~\ref{cor:integer}. For all other
$\alpha>0$, $s$ is not an integer, and the proof uses the whole
semigroup; compare Proposition~\ref{prop:compatibility}.
\end{remark}

\begin{remark}\label{rem:laplace}
The realization of Theorem~\ref{thm:realization} is related by the
Laplace transform $\mathcal Lf(p)=\int_0^\infty e^{-pt}f(t)\,dt$ to the
differential realization $v^a\mapsto y^a$, $u\mapsto hy^s\,d/dy$ of
$\As$ on $C^\infty((0,\infty))$ \cite[Eq.~(6)]{MSS2011},
\cite[Section~4]{MSS2012}. Let $\mathcal G$ be the space of all
$f\in\E$ whose derivatives have at most polynomial growth; $X$ and all
$I^r$ map it into itself. For $f\in\mathcal G$ and $p>0$, the
convolution theorem and integration by parts give
\[
 \mathcal L(I^rf)(p)=p^{-r}\,\mathcal Lf(p),\qquad
 \mathcal L(Xf)(p)=-\frac{d}{dp}\,\mathcal Lf(p),
\]
and with $y=p^{-\alpha}$ we have $p^{-\alpha a}=y^a$ and
$-d/dp=\alpha y^s\,d/dy$. Hence, on $\mathcal G$, the Laplace transform
carries \eqref{eq:IX} and~\eqref{eq:XI} to \eqref{eq:normal}
and~\eqref{eq:reverse} in the differential realization with $h=\alpha$.
These are special cases of the expansions of
Maier~\cite[Theorem~6.4]{Maier}. This correspondence holds only on
$\mathcal G$. Theorem~\ref{thm:L1} deals with larger function spaces,
and for derivatives this difference matters (Remark~\ref{rem:domain}).
The probabilistic formulas of Section~\ref{sec:kernel} are not obtained
in this way.
\end{remark}

\section{Coefficients, mixed products and special cases}\label{sec:special}

\begin{proposition}\label{prop:explicit}
Let $\alpha\in\R\setminus\{0\}$, $s=1+1/\alpha$, and $n\ge1$. Then
\begin{align}
 \prod_{i=0}^{n-1}\bigl(z+(\alpha+1)i\bigr)
   &=\sum_{k=1}^n S_{s;\alpha}(n,k)\,(z)_k,\label{eq:rising}\\
 S_{s;\alpha}(n,k)
   &=\frac1{k!}\sum_{j=0}^k(-1)^j\binom kj
       \prod_{i=0}^{n-1}\bigl((\alpha+1)i-j\bigr),\label{eq:explicit}
\end{align}
and $S_{s;\alpha}(n,k)$ is a polynomial in $\alpha$ of degree $n-k$
with nonnegative integer coefficients. Moreover, for $\alpha>0$ and
$\mu>-1$,
\begin{equation}\label{eq:powers-x}
 (I^\alpha X)^nx^\mu
   =\frac{\Gamma(\mu+1)}{\Gamma\bigl(\mu+1+(\alpha+1)n\bigr)}
    \prod_{i=0}^{n-1}\bigl(\mu+1+(\alpha+1)i\bigr)\,
    x^{\mu+(\alpha+1)n},
\end{equation}
and \eqref{eq:rising} with $z=\mu+1$ is the identity~\eqref{eq:IX}
applied to $x^\mu$.
\end{proposition}
\begin{proof}
Identity~\eqref{eq:rising} is Lemma~\ref{lem:connection} with
$h=\alpha$, since $hs=\alpha+1$ and $h(1-s)=-1$. Replacing $z$ by $-z$
turns its right-hand side into
$\sum_k(-1)^kS_{s;\alpha}(n,k)\,z(z-1)\cdots(z-k+1)$, and Newton's
forward-difference formula gives~\eqref{eq:explicit}. For $h=\alpha$
the recurrence~\eqref{eq:recurrence} has the multiplier
$\alpha(k+s(n-k))=n\alpha+(n-k)$, and this gives the statement about
polynomials. Finally, $I^rx^\nu=\Gamma(\nu+1)x^{\nu+r}/\Gamma(\nu+1+r)$
for $\nu>-1$ and $r\ge0$, so that
\[
 I^\alpha Xx^\nu=(\nu+1)\,
   \frac{\Gamma(\nu+1)}{\Gamma(\nu+\alpha+2)}\,x^{\nu+\alpha+1}.
\]
We iterate this with $\nu=\mu+(\alpha+1)i$, $i=0,\dots,n-1$; most gamma
factors cancel, and we obtain~\eqref{eq:powers-x}. The right-hand side
of~\eqref{eq:IX} maps $x^\mu$ to
$\sum_kS_{s;\alpha}(n,k)\Gamma(\mu+k+1)
x^{\mu+(\alpha+1)n}/\Gamma(\mu+1+(\alpha+1)n)$, and
$\Gamma(\mu+k+1)=(\mu+1)_k\,\Gamma(\mu+1)$.
\end{proof}

By~\eqref{eq:scaling}, the coefficients of the derivative
expansions \eqref{eq:DX} and~\eqref{eq:XD} are obtained
from~\eqref{eq:explicit} with $\alpha=-\beta$:
$S_{1-1/\beta;\beta}(n,k)=(-1)^{n-k}S_{1-1/\beta;-\beta}(n,k)$.

\Needspace{9\baselineskip}
\begin{corollary}\label{cor:special}
Let $n\ge1$.
\begin{enumerate}[\upshape(a)]
\item $(\alpha=1.)$ Let $V=I^1$ and
$c(m,j)=\dfrac{(m+j)!}{2^jj!\,(m-j)!}$. On $\Lloc$,
\[
 (XV)^n=\sum_{j=0}^{n-1}(-1)^jc(n-1,j)X^{n-j}V^{n+j},\qquad
 (VX)^n=\sum_{j=0}^{n-1}c(n-1,j)V^{n+j}X^{n-j}.
\]
\item $(\beta=1.)$ On $C^n((0,\infty))$,
\[
 (XD)^n=\sum_{k=1}^nS(n,k)X^kD^k,\qquad
 (DX)^n=\sum_{k=1}^n(-1)^{n-k}S(n,k)D^kX^k.
\]
\item $(\beta=\frac12.)$ For $f$ in the range of $I^{n/2}$ on $\Lloc$,
\[
 (XD^{1/2})^nf=\sum_{n/2\le k\le n}2^{k-n}B(n,k)\,X^kD^{k-n/2}f,
 \qquad B(n,k)=\frac{n!}{2^{n-k}(2k-n)!\,(n-k)!}.
\]
\end{enumerate}
\end{corollary}
\begin{proof}
(a) Here $s=2$, and $S_{2;1}(n,n-j)=c(n-1,j)$: by~\eqref{eq:recurrence}
this is equivalent to $c(n,j)=c(n-1,j)+(n+j-1)c(n-1,j-1)$, which can be
checked directly; see also~\cite{MSSMeromorphic,MSS2012}. Now apply
Theorem~\ref{thm:L1}(a) and~\eqref{eq:scaling}.
(b) Here $s=0$, and \eqref{eq:recurrence} with $h=1$ is the recurrence
of the Stirling numbers of the second kind; see
Remark~\ref{rem:domain} for the domain.
(c) Here $s=-1$ and
$S_{-1;1/2}(n,k)=2^{k-n}S_{-1;1}(n,k)$. The numbers $S_{-1;1}(n,k)$
satisfy $S(n+1,k)=S(n,k-1)+(2k-n)S(n,k)$, which is the recurrence of
the number $B(n,k)$ of partitions of an $n$-set into $k$ blocks of
size at most two: the new element is either a block by itself or it
joins one of the $2k-n$ singletons. Now apply~\eqref{eq:XD}, noting that
$(\beta-1)n+k=k-n/2$.
\end{proof}

The first formula in~(a) is Benmoussa's
expansion~\cite[Theorem~1]{BenmoussaJIS}; the numbers $c(m,j)$ are the
coefficients of the Bessel polynomials \cite[A001498]{OEIS}. Part~(b) is
Scherk's classical identity and its reversed form~\cite{MansourSchork},
and the $B(n,k)$ in~(c) are the Bessel numbers of the second
kind~\cite{MSS2012}. The numbers in~(c) also appear for a differential
operator. Since $[D,x^{a/2}]=\frac a2x^{a/2-1}$ for $a\in\Z$, the
assignment $v^a\mapsto x^{a/2}$, $u\mapsto D$ realizes $A_{-1,1/2}$ on
$C^\infty((0,\infty))$, and \eqref{eq:normal} gives
$(x^{1/2}D)^n=\sum_{n/2\le k\le n}2^{k-n}B(n,k)\,x^{k-n/2}D^k$. After
multiplication by $2^{\lfloor n/2\rfloor}$, these coefficients are the
Stirling numbers of order $1/2$ introduced by
Katugampola~\cite{Katugampola}, in the normalization of
Schork~\cite{Schork}. Schork identified them with a scaled subfamily of
the numbers~\eqref{eq:HS}.

Products with a fractional operator on each side of $X$ involve the
full family~\eqref{eq:HS}.

\begin{proposition}\label{prop:three}
Let $\gamma,\delta\in\R$, $\alpha=\gamma+\delta$, and $n\in\Nzero$.
On $\E$,
\begin{equation}\label{eq:three}
\begin{aligned}
 (I^\gamma XI^\delta)^n
  &=\sum_{k=0}^n S(n,k;\alpha+1,1,-\gamma)\,X^kI^{(\alpha+1)n-k}\\
  &=\sum_{k=0}^n(-1)^{n-k}S(n,k;\alpha+1,1,-\delta)\,
        I^{(\alpha+1)n-k}X^k.
\end{aligned}
\end{equation}
If $\gamma,\delta\ge0$, these identities hold on $\Lloc$.
\end{proposition}
\begin{proof}
By~\eqref{eq:group}, $I^\gamma XI^\delta=XI^\alpha-\gamma I^{\alpha+1}$
and, with $N=(\alpha+1)n$,
\[
 X^kI^{N-k}\bigl(XI^\alpha-\gamma I^{\alpha+1}\bigr)
   =X^{k+1}I^{N+\alpha-k}+(k-N-\gamma)\,X^kI^{N+\alpha+1-k}.
\]
Hence the coefficients $C(n,k)$ of the first expansion satisfy
$C(0,0)=1$ and
$C(n+1,k)=C(n,k-1)+\bigl(k-(\alpha+1)n-\gamma\bigr)C(n,k)$.
This is the recurrence
$S(n+1,k)=S(n,k-1)+(kb-na+r)S(n,k)$ of the numbers
$S(n,k)=S(n,k;a,b,r)$, which follows from~\eqref{eq:HS} after
multiplication by $z-na=(z-r-kb)+(kb-na+r)$. The second expansion is
obtained in the same way from
$I^\gamma XI^\delta=I^\alpha X+\delta I^{\alpha+1}$ and
multiplication on the left. If $\gamma,\delta\ge0$, only operators of
nonnegative order occur, and Lemma~\ref{lem:L1} applies.
\end{proof}

For $\gamma=0$ or $\delta=0$, \eqref{eq:three} contains \eqref{eq:XI}
and~\eqref{eq:IX} and their expansions in the opposite order. For
$\gamma=-\beta<0<\delta$, it expands $(D^\beta XI^\delta)^n$. For
$\delta=-\gamma=\beta$, it is the Leibniz rule for
$D^\beta X^nI^\beta=(I^{-\beta}XI^{\beta})^n$, since
$S(n,k;1,1,\beta)=\binom nk\beta(\beta-1)\cdots(\beta-n+k+1)$.

\section{A probabilistic formula}\label{sec:kernel}

The expansion~\eqref{eq:XI} has alternating signs, although
$(XI^\alpha)^n$ maps nonnegative functions to nonnegative functions. Its
kernel explains this and leads to a probabilistic description of the
coefficients. In this section, $\alpha>0$, $s=1+1/\alpha$, and $n\ge1$.
Let $(D_1,\dots,D_n)$ have the
symmetric Dirichlet distribution with parameter $\alpha$~\cite{NgTianTang},
that is, the density
$\Gamma(n\alpha)\Gamma(\alpha)^{-n}\prod_{i=1}^nd_i^{\alpha-1}$ on the
simplex $\{d_i>0,\ d_1+\dots+d_n=1\}$ with respect to
$dd_1\cdots dd_{n-1}$. Put $B_j=D_1+\dots+D_j$ for $1\le j\le n-1$, let
$\mathbb E$ denote expectation, and let $e_m$ be the $m$th elementary
symmetric polynomial.

\begin{theorem}\label{thm:kernel}
For $f\in\Lloc$ and almost every $x>0$,
\begin{equation}\label{eq:kernel}
 ((XI^\alpha)^nf)(x)=\int_0^xK_{n,\alpha}(x,t)f(t)\,dt,
 \quad
 K_{n,\alpha}(x,t)=\frac{x^n(x-t)^{n\alpha-1}}{\Gamma(n\alpha)}\,
   P_{n,\alpha}(1-t/x),
\end{equation}
where
\begin{equation}\label{eq:P}
 P_{n,\alpha}(y)=\sum_{m=0}^{n-1}
    \frac{S_{s;-\alpha}(n,n-m)}{(n\alpha)_m}\,y^m
   =\mathbb E\prod_{j=1}^{n-1}(1-yB_j).
\end{equation}
\end{theorem}
\begin{proof}
By~\eqref{eq:RL}, the operator $X^{n-m}I^{n\alpha+m}$ has the kernel
$x^{n-m}(x-t)^{n\alpha+m-1}/\Gamma(n\alpha+m)$, and
$\Gamma(n\alpha+m)=(n\alpha)_m\Gamma(n\alpha)$. Hence \eqref{eq:XI},
which holds on $\Lloc$ by Theorem~\ref{thm:L1}(a), gives
\eqref{eq:kernel} with the first expression for $P_{n,\alpha}$
in~\eqref{eq:P}. On the other hand, iterating~\eqref{eq:RL} and using
Tonelli's theorem, we obtain
$((XI^\alpha)^nf)(x)=\int_0^x\widetilde K(x,t)f(t)\,dt$ with
\[
 \widetilde K(x,t)=\frac x{\Gamma(\alpha)^n}
  \int_{t<z_{n-1}<\dots<z_1<x}\ \prod_{j=1}^{n-1}z_j\,
  \prod_{i=1}^n(z_{i-1}-z_i)^{\alpha-1}\,dz_1\cdots dz_{n-1},
\]
where $z_0=x$ and $z_n=t$. Put $y=(x-t)/x$ and
$d_i=(z_{i-1}-z_i)/(x-t)$. Then $d_i>0$, $d_1+\dots+d_n=1$,
$z_j=x\bigl(1-y(d_1+\dots+d_j)\bigr)$, and
$dz_1\cdots dz_{n-1}=(x-t)^{n-1}dd_1\cdots dd_{n-1}$, so that
\[
 \widetilde K(x,t)=\frac{x^n(x-t)^{n\alpha-1}}{\Gamma(n\alpha)}\,
   \mathbb E\prod_{j=1}^{n-1}(1-yB_j).
\]
Expanding the product, we see that $(XI^\alpha)^n$ is a linear
combination of the operators $X^{n-m}I^{n\alpha+m}$, $0\le m\le n-1$.
The coefficient of each of these operators is $(n\alpha)_m$ times the
corresponding coefficient of the polynomial
$\mathbb E\prod_j(1-yB_j)$. By the uniqueness of the coefficients
in~\eqref{eq:XI} (Corollary~\ref{cor:flat}), this polynomial is
$P_{n,\alpha}$.
\end{proof}

\begin{corollary}\label{cor:dirichlet}
\begin{enumerate}[\upshape(a)]
\item For $0\le m\le n-1$,
\begin{equation}\label{eq:moment}
 S_{s;\alpha}(n,n-m)=(n\alpha)_m\,\mathbb E\,e_m(B_1,\dots,B_{n-1}).
\end{equation}
Equivalently, for every real $s>1$, with Dirichlet parameter
$1/(s-1)$,
\begin{equation}\label{eq:prob}
 S_{s;1}(n,n-m)=\mathbb E\,e_m(B_1,\dots,B_{n-1})
     \prod_{i=0}^{m-1}\bigl(n+i(s-1)\bigr).
\end{equation}
\item As polynomials,
$P_{n,\alpha}(y)=(1-y)^{n-1}P_{n,\alpha}\bigl(-y/(1-y)\bigr)$.
Equivalently, on $\Lloc$,
\begin{equation}\label{eq:positive}
 (XI^\alpha)^n=\sum_{k=1}^nS_{s;\alpha}(n,k)\,
     XI^{(\alpha+1)n-k}X^{k-1},
\end{equation}
an expansion with positive coefficients.
\end{enumerate}
\end{corollary}
\begin{proof}
(a) Expanding the product in~\eqref{eq:P} gives
$P_{n,\alpha}(y)=\sum_m(-y)^m\,\mathbb E\,e_m(B_1,\dots,B_{n-1})$.
Comparing coefficients and using~\eqref{eq:scaling}, we
obtain~\eqref{eq:moment}. Then \eqref{eq:prob} follows from
$S_{s;\alpha}(n,n-m)=\alpha^mS_{s;1}(n,n-m)$ and
$\alpha^{-m}(n\alpha)_m=\prod_{i<m}(n+i/\alpha)$, where $1/\alpha=s-1$.
(b) The vector $(D_n,\dots,D_1)$ has the same distribution as
$(D_1,\dots,D_n)$, and this reversal replaces $(B_1,\dots,B_{n-1})$ by
$(1-B_{n-1},\dots,1-B_1)$. Hence
\[
 \mathbb E\prod_{j=1}^{n-1}(1-yB_j)
   =\mathbb E\prod_{j=1}^{n-1}(1-y+yB_j)
   =(1-y)^{n-1}\,\mathbb E\prod_{j=1}^{n-1}\Bigl(1+\frac y{1-y}B_j\Bigr),
\]
which is the first statement. With $y=(x-t)/x$ it reads
$x^nP_{n,\alpha}(y)=x\,t^{n-1}P_{n,\alpha}\bigl(-(x-t)/t\bigr)$, so
that, by \eqref{eq:P} and~\eqref{eq:scaling},
\[
 K_{n,\alpha}(x,t)=\sum_{m=0}^{n-1}S_{s;\alpha}(n,n-m)\,
   \frac{x\,t^{n-1-m}(x-t)^{n\alpha+m-1}}{\Gamma(n\alpha+m)}.
\]
The $m$th term is the kernel of $XI^{n\alpha+m}X^{n-1-m}$, which
gives~\eqref{eq:positive} with $k=n-m$; the computation can be reversed.
\end{proof}

Identity~\eqref{eq:positive}, composed with $X$ on the right,
is~\eqref{eq:IX} multiplied by $X$ on the left, because
$X(I^\alpha X)^n=(XI^\alpha)^nX$. Thus the reversal symmetry of the
Dirichlet distribution corresponds to switching between the two orders
\eqref{eq:IX} and~\eqref{eq:XI}. The kernel also gives simple bounds.
Since $0<B_j<1$ almost surely, we have
$(1-y)^{n-1}\le P_{n,\alpha}(y)\le1$ for $0\le y\le1$. With
$y=(x-t)/x$, the functions $x\,t^{n-1}$, $x^nP_{n,\alpha}(y)$ and $x^n$,
multiplied by
$(x-t)^{n\alpha-1}/\Gamma(n\alpha)$, are the kernels of
$XI^{n\alpha}X^{n-1}$, $(XI^\alpha)^n$ and $X^nI^{n\alpha}$. Hence, for
every nonnegative $f\in\Lloc$, almost everywhere,
\begin{equation}\label{eq:sandwich}
 XI^{n\alpha}X^{n-1}f\le(XI^\alpha)^nf\le X^nI^{n\alpha}f.
\end{equation}

\subsection{An urn model}
In the proof of Theorem~\ref{thm:kernel}, the Dirichlet density comes
from the factors $(z_{i-1}-z_i)^{\alpha-1}$ of the iterated integral. By
de Finetti's theorem, the Dirichlet distribution also appears in
P\'olya urns~\cite{JohnsonKotz}, and this leads to a description of the
coefficients by urns. For a sequence $i=(i_1,\dots,i_m)$ of colours
$i_t\in\{1,\dots,n\}$, let $c_r(i)$ be the number of $t$ with $i_t=r$,
and define
\begin{equation}\label{eq:weight}
 w_s(i)=\prod_{r=1}^n\ \prod_{q=0}^{c_r(i)-1}\bigl(1+q(s-1)\bigr).
\end{equation}

\begin{proposition}\label{prop:urn}
Let $n\ge1$ and $0\le m\le n-1$. For every $s\in\R$,
\begin{equation}\label{eq:urn}
\begin{aligned}
 S_{s;1}(n,n-m)&=\sum_{1\le j_1<\dots<j_m\le n-1}\ \prod_{t=1}^m
   \bigl(j_t+(t-1)(s-1)\bigr)\\
 &=\sum_{1\le j_1<\dots<j_m\le n-1}\ \
 \sum_{1\le i_1\le j_1,\,\dots,\,1\le i_m\le j_m}w_s(i).
\end{aligned}
\end{equation}
Now let $s\ge1$, or $s=1-1/K$ with $K\in\N$, and let $I_1,I_2,\dots$ be
random colours in $\{1,\dots,n\}$ such that
\begin{equation}\label{eq:urn-rule}
 \Prob(I_{t+1}=r\mid I_1,\dots,I_t)=\frac{1+(s-1)c_r}{n+(s-1)t},
\end{equation}
where $c_r$ is the number of $t'\le t$ with $I_{t'}=r$. Then
\begin{equation}\label{eq:urn-prob}
 S_{s;1}(n,n-m)=\prod_{t=0}^{m-1}\bigl(n+t(s-1)\bigr)
   \sum_{1\le j_1<\dots<j_m\le n-1}\Prob(I_1\le j_1,\dots,I_m\le j_m).
\end{equation}
\end{proposition}

For $s\ge1$, \eqref{eq:urn-rule} describes a P\'olya urn. At the start
the urn contains one ball of each of the $n$ colours. After each draw,
the ball is put back together with $s-1$ more balls of the same colour
(if $s-1$ is not an integer, we use weights instead of balls). For
$s=1-1/K$, \eqref{eq:urn-rule} reads
$\Prob(I_{t+1}=r\mid\dots)=(K-c_r)/(nK-t)$; that is, the balls are drawn
without replacement from an urn with $K$ balls of each colour. By
Theorem~\ref{thm:L1}, the values $s=1+1/\alpha$ correspond to the
fractional integrals $I^\alpha$, and the values $s=1-1/K$ correspond to
the derivatives $D^K$ of integer order.

\begin{proof}
Put $T(n,m)=S_{s;1}(n,n-m)$. By \eqref{eq:recurrence} with $h=1$ and
$k=n+1-m$,
\[
 T(n+1,m)=T(n,m)+\bigl(n+(m-1)(s-1)\bigr)\,T(n,m-1)\qquad(1\le m\le n),
\]
where $T(n,0)=1$, and $T(n,m)=0$ for $m\ge n\ge1$. Applying this
recurrence repeatedly in $n$ gives the first expression
in~\eqref{eq:urn}. For the second, fix $j$ and sum over $i_m$ first.
Since $i_1,\dots,i_{m-1}\le j_{m-1}<j_m$, the colours
$r\le j_m$ have the weights $1+c_r(s-1)$, where $c_r$ is the number of
$t<m$ with $i_t=r$ and $\sum_rc_r=m-1$. Hence the sum over $i_m$
multiplies $w_s(i_1,\dots,i_{m-1})$ by $j_m+(m-1)(s-1)$, and induction on
$m$ gives the product.

For $s\ge1$ the right side of \eqref{eq:urn-rule} is nonnegative, and its
sum over $r$ is $1$ because $\sum_rc_r=t$. For $s=1-1/K$ it equals
$(K-c_r)/(nK-t)$, with $c_r\le K$ and $nK-t>0$ for $t<m\le n-1$. In both
cases the denominators $n+t(s-1)$, $t<m$, are positive. When we
multiply the conditional probabilities, the draws of colour $r$ give the
factors $1,\,1+(s-1),\,\dots,\,1+(c_r(i)-1)(s-1)$, so
$\Prob(I_1=i_1,\dots,I_m=i_m)=w_s(i)/\prod_{t<m}(n+t(s-1))$. Summing over
the $i$ with $i_t\le j_t$ and over $j$, and using \eqref{eq:urn}, we
obtain~\eqref{eq:urn-prob}.
\end{proof}

For $s>1$, \eqref{eq:urn-prob} and \eqref{eq:prob} are two forms of the
same fact. A P\'olya urn is a mixture of independent draws whose colour
probabilities form a Dirichlet vector $D$ with parameter
$1/(s-1)$~\cite{JohnsonKotz}, and, given $D$, the event
$\{I_t\le j_t\ \text{for all }t\}$ has probability
$B_{j_1}\cdots B_{j_m}$. Hence, with $\alpha=1/(s-1)$,
\[
 P_{n,\alpha}(y)=\sum_{m=0}^{n-1}(-y)^m
   \sum_{1\le j_1<\dots<j_m\le n-1}\Prob(I_1\le j_1,\dots,I_m\le j_m),
\]
so the coefficients of the kernel in Theorem~\ref{thm:kernel} are urn
probabilities. Also, \eqref{eq:prob} follows from
Proposition~\ref{prop:urn}, independently of Theorem~\ref{thm:kernel}. As
$s\downarrow1$, the urn adds fewer and fewer balls, so the draws become
independent and uniform, and $B_j$ concentrates at $\mathbb EB_j=j/n$.
So \eqref{eq:prob} is a random version of the classical formula in
Corollary~\ref{cor:urn-special}(a) below, with the points $j/n$ replaced
by the random points $B_j$.

\begin{corollary}\label{cor:urn-special}
Let $n\ge2$ and $1\le m\le n-1$.
\begin{enumerate}[\upshape(a)]
\item \textup{($s=1$, independent uniform draws.)}
$S_{1;1}(n,n-m)=e_m(1,2,\dots,n-1)$, the unsigned Stirling numbers of the
first kind.
\item \textup{($s=0$, drawing without replacement, $K=1$.)}
\[
 S_{0;1}(n,n-m)=\sum_{1\le j_1<\dots<j_m\le n-1}\ \prod_{t=1}^m(j_t-t+1)
 =h_m(1,2,\dots,n-m),
\]
the Stirling numbers $S(n,n-m)$ of the second kind; here $h_m$ is the
complete homogeneous symmetric polynomial.
\item \textup{($s=2$, a P\'olya urn adding one ball.)}
$S_{2;1}(n,n-m)=\sum_j\sum_i\prod_rc_r(i)!=c(n-1,m)$, the Bessel numbers
of Corollary~\ref{cor:special}(a).
\item As a polynomial in $s$, $S_{s;1}(n,n-m)$ has degree $m-1$ and
leading coefficient $(m-1)!\binom n{m+1}$; for example,
$S_{s;1}(5,1)=(s+1)(2s+1)(3s+1)$.
\end{enumerate}
\end{corollary}
\begin{proof}
(a) For $s=1$ the product in~\eqref{eq:urn} is $j_1\cdots j_m$, and
\eqref{eq:recurrence} with $h=1$ is the recurrence
$S(n+1,k)=S(n,k-1)+nS(n,k)$ of the unsigned Stirling numbers of the
first kind~\cite[Section~2]{MSS2011}.
(b) For $s=0$ the product is $\prod_t(j_t-t+1)$. In the urn, $j_t-t+1$
is the number of colours $r\le j_t$ that have not been drawn before. With
$l_t=j_t-t+1$ the sum runs over $1\le l_1\le\dots\le l_m\le n-m$, which
is $h_m(1,\dots,n-m)$. For $s=0$ the numbers are the Stirling numbers of
the second kind, as in the proof of Corollary~\ref{cor:special}(b).
(c) Here $w_2(i)=\prod_r\prod_{q<c_r(i)}(1+q)=\prod_rc_r(i)!$, and
$S_{2;1}(n,n-m)=c(n-1,m)$ by the proof of Corollary~\ref{cor:special}(a).
(d) In the first sum of~\eqref{eq:urn}, the term of $j$ is
$j_1\prod_{t=2}^m\bigl(j_t+(t-1)(s-1)\bigr)$, a polynomial in $s$ of
degree $m-1$ with leading coefficient $(m-1)!\,j_1$. Moreover
$\sum_jj_1=\binom n{m+1}$, because the sets $\{r-1,j_1,\dots,j_m\}$ with
$1\le r\le j_1$ are exactly the $(m+1)$-element subsets of
$\{0,1,\dots,n-1\}$. The example follows from \eqref{eq:recurrence},
which gives $S_{s;1}(n+1,1)=(1+(n-1)s)S_{s;1}(n,1)$.
\end{proof}

\begin{remark}\label{rem:pitman}
The same numbers occur in a different urn model. By the identification
after~\eqref{eq:HS} and the scaling
$S(n,k;ca,cb,cr)=c^{n-k}S(n,k;a,b,r)$, which follows from~\eqref{eq:HS},
\[
 S_{s;1}(n,k)=s^{n-k}S(n,k;-1,-\tau,0),\qquad\tau=1-1/s\quad(s\ne0).
\]
Let $s\ge1$, that is, $0\le\tau<1$, and let $\theta>-\tau$. Then the
numbers on the right give the distribution of the number $K_n$ of blocks
of the Ewens--Pitman partition of $\{1,\dots,n\}$ with parameters
$(\tau,\theta)$. This random partition is generated by the Chinese
restaurant process~\cite[Sections~3.1--3.2]{Pitman}, and
\[
 \Prob(K_n=k)=\frac{\prod_{i=1}^{k-1}(\theta+i\tau)}{(\theta+1)_{n-1}}\,
   S(n,k;-1,-\tau,0).
\]
This model counts the blocks, while~\eqref{eq:urn-prob} is about
threshold events for a fixed number $n$ of colours. Urn models for other
families of generalized Stirling numbers appear in~\cite{JKP}.
\end{remark}

\begin{remark}\label{rem:limits}
For $s=2$, that is, $\alpha=1$, the Dirichlet distribution is uniform on
the simplex, and $(B_1,\dots,B_{n-1})$ has the law of the order
statistics of $n-1$ independent uniform variables on
$(0,1)$~\cite{DavidNagaraja}. Since the product in~\eqref{eq:P} is
symmetric in the $B_j$, we get $P_{n,1}(y)=(1-y/2)^{n-1}$ and
$K_{n,1}(x,t)=x(x^2-t^2)^{n-1}/(2^{n-1}(n-1)!)$. In general $P_{n,\alpha}$
is not a power of a linear polynomial; for example,
$P_{3,\alpha}(y)=1-y+\frac{2\alpha+1}{3(3\alpha+1)}y^2$.
\end{remark}

\subsection{\texorpdfstring{Words in $X$ and fractional integrals}{Words in X and fractional integrals}}
The method of Theorem~\ref{thm:kernel} works for arbitrary words. Let
$n\ge1$, $\alpha_1,\dots,\alpha_n>0$, $\lambda_1,\dots,\lambda_n\in\Nzero$,
and
\[
 w=X^{\lambda_n}I^{\alpha_n}X^{\lambda_{n-1}}I^{\alpha_{n-1}}\cdots
   X^{\lambda_1}I^{\alpha_1},\qquad
 \mathsf A=\sum_{i=1}^n\alpha_i,\qquad\Lambda=\sum_{i=1}^n\lambda_i.
\]
In this subsection, $(D_1,\dots,D_n)$ is a Dirichlet vector with
parameters $(\alpha_n,\dots,\alpha_1)$, so that
\[
 \mathbb E\prod_{r=1}^nD_r^{c_r}
   =\frac{\prod_{r=1}^n(\alpha_{n+1-r})_{c_r}}{(\mathsf A)_{c_1+\dots+c_n}}
   \qquad(c_r\in\Nzero)
\]
\cite{NgTianTang}, and again $B_j=D_1+\dots+D_j$. Put
\begin{equation}\label{eq:Qw}
 Q_w(y)=\mathbb E\prod_{j=1}^{n-1}(1-yB_j)^{\lambda_{n-j}}
   =\sum_{m=0}^{\Lambda-\lambda_n}(-1)^mq_my^m.
\end{equation}

\begin{theorem}\label{thm:words}
In this notation, $q_m>0$ for $0\le m\le\Lambda-\lambda_n$, and on $\Lloc$
\begin{equation}\label{eq:words}
 w=\sum_{m=0}^{\Lambda-\lambda_n}(-1)^m(\mathsf A)_m\,q_m\,
   X^{\Lambda-m}I^{\mathsf A+m}.
\end{equation}
Equivalently, for $f\in\Lloc$ and almost every $x>0$,
\[
 (wf)(x)=\int_0^x\frac{x^\Lambda(x-t)^{\mathsf A-1}}{\Gamma(\mathsf A)}\,
   Q_w\Bigl(\frac{x-t}x\Bigr)f(t)\,dt.
\]
Explicitly,
\[
 q_m=\sum_{m_1+\dots+m_{n-1}=m}\ \prod_{j=1}^{n-1}\binom{\lambda_{n-j}}{m_j}\
   \mathbb E\prod_{j=1}^{n-1}B_j^{m_j},
\]
and these mixed moments are finite sums of the moments of the $D_r$.
\end{theorem}
\begin{proof}
Iterating~\eqref{eq:RL} and using Tonelli's theorem, we obtain
$(wf)(x)=\int_0^x\widetilde K(x,t)f(t)\,dt$ with
\[
 \widetilde K(x,t)=x^{\lambda_n}\int_{t<z_{n-1}<\dots<z_1<x}\
   \prod_{j=1}^{n-1}z_j^{\lambda_{n-j}}
   \prod_{i=1}^n\frac{(z_{i-1}-z_i)^{\alpha_{n+1-i}-1}}{\Gamma(\alpha_{n+1-i})}
   \,dz_1\cdots dz_{n-1},
\]
where $z_0=x$ and $z_n=t$. The substitution in the proof of
Theorem~\ref{thm:kernel} turns this into
$x^\Lambda(x-t)^{\mathsf A-1}\Gamma(\mathsf A)^{-1}Q_w((x-t)/x)$. The
function $x^{\Lambda-m}(x-t)^{\mathsf A+m-1}/\Gamma(\mathsf A)$ is
$(\mathsf A)_m$ times the kernel of $X^{\Lambda-m}I^{\mathsf A+m}$, so
expanding $Q_w$ gives~\eqref{eq:words}. The formula for $q_m$ follows
from the binomial theorem. For $m\le\Lambda-\lambda_n
=\sum_{j=1}^{n-1}\lambda_{n-j}$ the sum that defines $q_m$ is not empty,
and each term is positive because $B_j>0$ almost surely.
\end{proof}

The urn model also extends to words. Consider a P\'olya urn in which
colour $r$ starts with weight $\alpha_{n+1-r}$ and gains weight $1$ each
time it is drawn. By de Finetti's theorem, given $(D_1,\dots,D_n)$ the
draws $I_1,I_2,\dots$ are independent with $\Prob(I_t=r)=D_r$
\cite{JohnsonKotz}. Hence
$\mathbb E\,B_{k_1}\cdots B_{k_M}=\Prob(I_1\le k_1,\dots,I_M\le k_M)$ for
all $k_1,\dots,k_M$, and each $q_m$ is a sum of such probabilities. For
$\alpha_r=\alpha$ this is the urn~\eqref{eq:urn-rule} with
$s=1+1/\alpha$.

\begin{corollary}\label{cor:power-words}
Let $\lambda\in\N$, $\alpha>0$ and $n\ge1$, and let the $B_j$ come from the
symmetric Dirichlet distribution with parameter $\alpha$. Define
$\mathfrak a^{\lambda,\alpha}_{n,m}$ as $(n\alpha)_m$ times the
coefficient of $y^m$ in $\mathbb E\prod_{j=1}^{n-1}(1+yB_j)^\lambda$. Then
$\mathfrak a^{\lambda,\alpha}_{n,m}>0$ for $0\le m\le\lambda(n-1)$, and
on $\Lloc$
\begin{align}
 (X^\lambda I^\alpha)^n&=\sum_{m=0}^{\lambda(n-1)}(-1)^m
   \mathfrak a^{\lambda,\alpha}_{n,m}\,X^{\lambda n-m}I^{\alpha n+m},
   \label{eq:XlI}\\
 (I^\alpha X^\lambda)^n&=\sum_{m=0}^{\lambda(n-1)}
   \mathfrak a^{\lambda,\alpha}_{n,m}\,I^{\alpha n+m}X^{\lambda n-m}.
   \label{eq:IXl}
\end{align}
These numbers satisfy $\mathfrak a^{\lambda,\alpha}_{1,0}=1$ and
\begin{equation}\label{eq:wordrec}
 \mathfrak a^{\lambda,\alpha}_{n+1,m}=\sum_{j}\binom{\lambda n-j}{m-j}
   (\alpha)_{m-j}\,\mathfrak a^{\lambda,\alpha}_{n,j},
\end{equation}
and $\mathfrak a^{1,\alpha}_{n,m}=\alpha^mS_{1+1/\alpha;1}(n,n-m)$.
\end{corollary}
\begin{proof}
The expansion~\eqref{eq:XlI} is Theorem~\ref{thm:words} with
$\lambda_i=\lambda$ and $\alpha_i=\alpha$. By induction on $p$,
Lemma~\ref{lem:L1}(b) gives
\[
 I^\alpha X^p=\sum_{i=0}^p(-1)^i\binom pi(\alpha)_i\,X^{p-i}I^{\alpha+i}.
\]
Write $(X^\lambda I^\alpha)^{n+1}=X^\lambda I^\alpha(X^\lambda I^\alpha)^n$,
expand each $I^\alpha X^{\lambda n-j}$ in this way, and collect the terms
with $X^{\lambda(n+1)-m}I^{\alpha(n+1)+m}$. This gives~\eqref{eq:wordrec},
because these operators are linearly independent
(Corollary~\ref{cor:flat}). For $\lambda=1$, compare~\eqref{eq:XlI}
with~\eqref{eq:XI} and use~\eqref{eq:scaling}. Finally,
$X^\lambda(I^\alpha X^\lambda)^n=(X^\lambda I^\alpha)^nX^\lambda$, so the
kernel of $(I^\alpha X^\lambda)^n$ is $(t/x)^\lambda$ times that of
$(X^\lambda I^\alpha)^n$. As in the proof of
Corollary~\ref{cor:dirichlet}(b),
$\mathbb E\prod_j(1-yB_j)^\lambda=(1-y)^{\lambda(n-1)}\,
\mathbb E\prod_j(1+y'B_j)^\lambda$ with $y'=y/(1-y)=(x-t)/t$. Hence this
kernel is
$t^{\lambda n}(x-t)^{n\alpha-1}\Gamma(n\alpha)^{-1}\,
\mathbb E\prod_j(1+y'B_j)^\lambda$, and expanding in powers of $y'$
gives~\eqref{eq:IXl}.
\end{proof}

For an integer order $\alpha=\delta$ we have
$\binom{\lambda n-j}{m-j}(\delta)_{m-j}
=\frac{(\lambda n-j)!}{(\lambda n-m)!}\binom{m-j+\delta-1}{\delta-1}$, so
\eqref{eq:wordrec} is the recurrence found by Benmoussa for
$(x^\lambda I^\delta)^n$~\cite[Section~3]{BenmoussaWords}.
Theorem~\ref{thm:words} gives these numbers in closed form, shows that
they are positive, and extends them to real orders. For the Volterra
operator the closed form is especially simple.

\begin{corollary}\label{cor:volterra-words}
Let $\lambda\in\N$ and $V=I^1$. On $\Lloc$,
\[
 (X^\lambda V)^n=\sum_{m=0}^{\lambda(n-1)}(-1)^m(n)_m\,g^{(\lambda)}_{n,m}\,
   X^{\lambda n-m}V^{n+m},\qquad
 \sum_mg^{(\lambda)}_{n,m}y^m
   =\Bigl(\frac{(1+y)^{\lambda+1}-1}{(\lambda+1)y}\Bigr)^{n-1}.
\]
\end{corollary}
\begin{proof}
For $\alpha=1$, $(B_1,\dots,B_{n-1})$ has the law of the order statistics
of $n-1$ independent uniform variables $U_1,\dots,U_{n-1}$ on
$(0,1)$~\cite{DavidNagaraja}. Since $\prod_j(1+yB_j)^\lambda$ is
symmetric in the $B_j$, its expectation is
$\bigl(\mathbb E(1+yU_1)^\lambda\bigr)^{n-1}$, and
$\mathbb E(1+yU_1)^\lambda=((1+y)^{\lambda+1}-1)/((\lambda+1)y)$. Now
apply Corollary~\ref{cor:power-words} with $\alpha=1$.
\end{proof}

For $\lambda=1$ we get $g^{(1)}_{n,m}=\binom{n-1}m2^{-m}$ and
$(n)_mg^{(1)}_{n,m}=c(n-1,m)$, which is Corollary~\ref{cor:special}(a)
again.

\section{Concluding remarks}\label{sec:concluding}

The range of $I^{n\beta}$ in Theorem~\ref{thm:L1}(b) is a sufficient
domain. It would be interesting to describe all functions for which
\eqref{eq:DX} and~\eqref{eq:XD} hold, and the correction terms that
appear in general. Remark~\ref{rem:domain} suggests that these terms
depend on the behaviour of $f$ at the origin. The urn interpretation
of Proposition~\ref{prop:urn} covers $s\ge1$ and $s=1-1/K$, that is,
fractional integrals of every order and derivatives of integer order.
For derivatives of non-integer order, that is, for $s<1$ with
$1/(1-s)\notin\N$, some weights $w_s(i)$ are negative (for $s=-1$, a
colour drawn twice gives $1-2=-1$). A probabilistic interpretation
of~\eqref{eq:urn} in this range would be welcome. Words that contain
fractional derivatives need conditions at the origin, as in
Theorem~\ref{thm:L1}(b), and the kernel method of
Theorem~\ref{thm:words} does not apply to them directly. Finally, similar
questions arise for the Caputo derivative, whose commutator with $X$
contains an initial-value term, and for the Hadamard and
Erd\'elyi--Kober operators.

\textbf{Funding.} This work was supported by the National University of
Mongolia.

\end{document}